\documentclass[reprint,amsmath,amssymb,floatfix,aps]{revtex4-1}
\usepackage{graphicx}% complex graphics
\usepackage{bm}% bold math
\usepackage{color}
\usepackage{amsmath}
\usepackage{amssymb}

\newcommand{\bs}{\boldsymbol}

\begin{document}
\title{Theory of Plasmon-Enhanced Metal Photoluminescence}

\author{Tigran V. Shahbazyan}
%\email{shahbazyan@jsums.edu}
\affiliation{Department of Physics, Jackson State University, Jackson, MS 39217, USA}

\begin{abstract}
Metal photoluminescence (MPL) originates from radiative recombination of photoexcited core holes and conduction band electrons. In metal nanostructures, MPL is enhanced due to the surface plasmon local field effect. We identify another essential process in plasmon-assisted MPL---excitation of Auger plasmons by core holes---that hinders MPL from small nanostructures. We develop a microscopic theory of plasmon-enhanced MPL that incorporates both plasmon-assisted enhancement and suppression mechanisms and derive the enhancement factor for MPL quantum efficiency. Our numerical calculations of MPL from Au nanoparticles are in excellent agreement with the experiment.
\end{abstract}

\maketitle

\section{Introduction}

Since its discovery by Mooradian in 1969 \cite{mooradian-prl69}, photoluminescence of noble metals and, later, metal nanostructures has attracted increasing interest \cite{apell-ps88,boyd-prb86,wilcoxon-jcp98,elsayed-cpl00,elsayed-jpcb03,novotny-prb03,feldmann-prb04,fourkas-nl05,wang-pnas05,bouhelier-prl05,imura-jpcb05,elsayed-prb05,schuck-prl05,eichelbaum-nanotech07,durr-nl07,wang-nl07,potma-jpcc08,park-oe08,imura-jpcc09,wang-oe09,biagioni-prb09,loumaigne-nl10,wissert-nl10,link-jpcc11,hulst-nl11,biagioni-nl12}. The underlying mechanism of metal photoluminescence (MPL) is radiative recombination of photoexcited d-band holes and upper-lying  s-band electrons---a process strongly suppressed in the bulk by a plethora of nonradiative transitions in noble metals \cite{apell-ps88}.  In confined metal structures,  MPL is considerably more efficient due to plasmon enhancement of radiative transitions \cite{boyd-prb86}. During the past decade,  a highly efficient single-photon and multiphoton-absorption-induced MPL was reported from various plasmonic systems including spherical nanoparticles (NP) \cite{wilcoxon-jcp98,novotny-prb03,feldmann-prb04,fourkas-nl05,eichelbaum-nanotech07,biagioni-prb09,loumaigne-nl10,wissert-nl10,hulst-nl11,biagioni-nl12}, nanorods \cite{elsayed-cpl00,elsayed-jpcb03,wang-pnas05,bouhelier-prl05,imura-jpcb05,elsayed-prb05,durr-nl07,imura-jpcc09,wang-oe09,link-jpcc11}, nanoshells \cite{park-oe08}, nanowires \cite{wang-nl07,potma-jpcc08}, and bowtie antennas \cite{schuck-prl05}. 
This continuing interest in plasmon-enhanced MPL is fueled, in large part, by its promising applications, e.g., in imaging of blood vessels \cite{wang-pnas05} and cancer cells \cite{durr-nl07} and optical recording \cite{zijlstra-nature09}.

\begin{figure}[bt]
  \begin{center}
%  \centering
  \includegraphics[width=0.95\columnwidth]{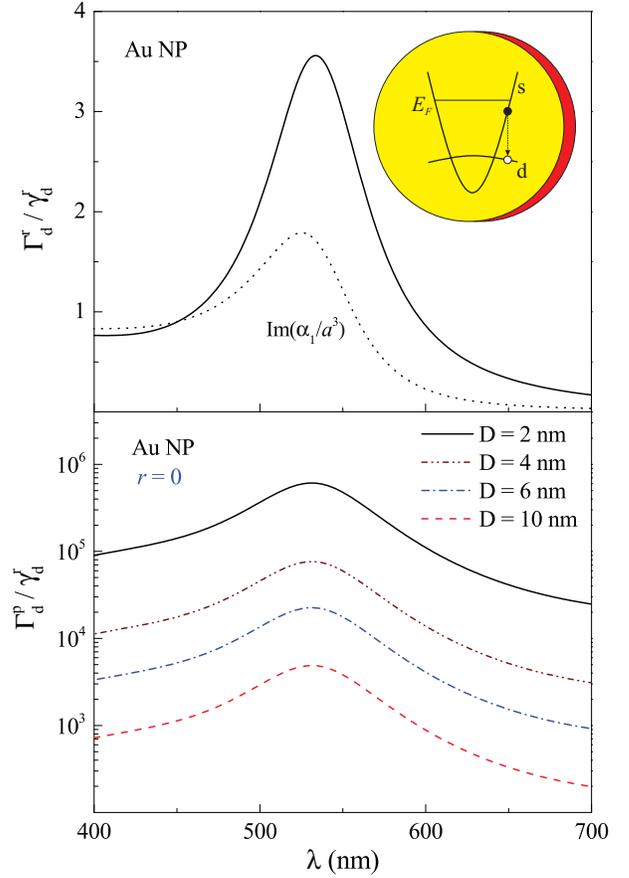}
  \end{center}
\caption{\label{fig:fig1} (a) Plasmon enhancement of d-hole radiative recombination for Au NP in water (solid line) and normalized absorption (dashed line). (b) Auger plasmon excitation rate by a  d-hole localized at the NP center. }
  \end{figure}

In this Letter, we present a microscopic theory of MPL from plasmonic nanostructures. To highlight the issue at hand, consider radiative recombination of a d-hole and an s-electron inside a spherical NP of radius $a$ much smaller than the radiation wavelength $\lambda$. Since d-holes in noble metals are strongly localized, this process can be viewed as radiation of a dipole located at the d-hole position ${\bf r}$. Radiative decay rate of such a dipole, $\Gamma_{d}^{r}$, is enhanced relative to free-space radiative decay rate, $\gamma_{d}^{r}$, by the local field factor \cite{boyd-prb86}
\begin{equation}
\label{rad}
R\equiv\dfrac{\Gamma_{d}^{r}}{\gamma_{d}^{r}}
%=\left |1-\frac{\alpha_{1}(\omega)}{a^{3}}\right |^{2}
=\left |\frac{3\epsilon_{0}}{\epsilon(\omega)+2\epsilon_{0}}\right |^{2}, 
\end{equation}
where $\epsilon(\omega)$ and $\epsilon_{0}$ are, respectively, metal and outside medium dielectric functions. The factor (\ref{rad}) has a peak when radiation frequency $\omega$ is close to surface plasmon frequency; the latter is defined as the pole in NP dipole polarizability, $\alpha_{1}=a^{3}(\epsilon-\epsilon_{0})/(\epsilon+2\epsilon_{0})$ [see Fig.~\ref{fig:fig1}(a)]. Importantly, Eq.~(\ref{rad}) depends neither on NP radius $a$ nor on d-hole position $r$, thus implying universal MPL enhancement within wide NP size range, $a\ll \lambda$.

%  %%%%%%%%%%%%%%%%%%%%%%%%%%%%%%%%%%
%  %%%%%%%%%%%%%%%%%%%%%%%%%%%%%%%%%%

However, the experiment points to a more complicated picture. Systematic studies of MPL size dependence for spherical NPs \cite{feldmann-prb04,fourkas-nl05} revealed a strong emission reduction, by a factor 3--5, for small NP with diameters $D\lesssim 3$ nm. Furthermore, MPL from larger NPs was reduced as well \cite{novotny-prb03,fourkas-nl05}, and the brightest NPs  observed were those with $D\sim 60$ nm \cite{fourkas-nl05}. While for large NPs, plasmon enhancement is expected to weaken due to the retardation effects, its decrease for smaller NPs indicates that another mechanism, sensitive to NP size, is at work. Here, we identify this mechanism as \textit{excitation of Auger plasmons by d-holes}.

Namely, a localized d-hole can undergo a  highly efficient \textit{nonradiative} recombination with an s-electron   accompanied by excitation of surface plasmons with \textit{high} angular momenta $l$. This process is analogous to the traditional d-hole Auger scattering to s-band with a Fermi sea electron--hole pair being excited except now collective, rather than single-particle, excitations populate the final state. Importantly, the Auger plasmon excitation rate, $\Gamma_{d}^{p}$, strongly depends on the proximity of d-hole position $r$ to the metal boundary. For example, excitation of high-$l$ Auger plasmons by a d-hole located near the center of a spherical NP ($r/a\ll 1$) is suppressed by symmetry and only the dipole ($l=1$) plasmon is excited with the rate \cite{shahbazyan-prl98,shahbazyan-prb99,shahbazyan-cp00} $\Gamma_{d}^{p}\propto a^{-3}$; but for a d-hole near NP boundary ($r/a\sim 1$), high-$l$  plasmons are efficiently excited causing $\Gamma_{d}^{p}$ increase by several orders of magnitude.  In small NPs, where even dipole Auger plasmons are efficiently excited \cite{shahbazyan-prl98,shahbazyan-prb99,shahbazyan-cp00},  the plasmon-assisted nonradiative decay rate $\Gamma_{d}^{p}$ can exceed (on average) the bulk  decay rate $\gamma_{d}^{nr}$. This  leads to the \textit{size-dependent reduction} of plasmon-enhanced MPL. In fact, d-hole radiative recombination  is quenched near NP boundary, and so the light emanates mainly from NP  central  region. Below we derive general expressions for MPL enhancement factor applicable to arbitrary metal nanostructures and present explicit formulas for a spherical NP. Our numerical calculations of MPL quantum efficiency (QE) for Au NP indicate that MPL from small NPs is significantly reduced and that brightest NPs are those with $D\approx 60$ nm, in excellent agreement with the experiment \cite{feldmann-prb04,fourkas-nl05}.

Before we proceed, let us compare classical and microscopic descriptions of MPL enhancement mechanism. Within the classical approach \cite{boyd-prb86}, the radiating dipole comprised of a d-hole and an s-electron is enhanced by the surface plasmon local field. Within the microscopic approach \cite{feldmann-prb04}, MPL enhancement is caused by the appearance of plasmon-mediated radiative decay channel: the d-hole recombines nonradiatively by exciting a surface plasmon \cite{shahbazyan-prl98,shahbazyan-prb99,shahbazyan-cp00} which then decays radiatively, leaving a photon in the final state. Within Fermi golden  rule, the photon emission rate is obtained by incorporating plasmons as intermediate states of the transition matrix element. Naturally, both approaches yield the same enhancement factor (\ref{rad}) for MPL intensity. However, excitation of high-$l$ Auger plasmons with lifetime  of $\sim$10 fs \cite{feldmann-prl98}  by localized d-holes provides a new nonradiative decay channel that can \textit{suppress} MPL from small plasmonic structures. Below, we incorporate Auger plasmons excitation processes within a microscopic model for MPL.

\section{Theory}

To specify the scope of our theory, let us elucidate the relevant processes  within the standard  three stage MPL model \cite{boyd-prb86,apell-ps88}. Initially, the incident light excites a nonequilibrium d-hole population which, during the second stage, undergoes fast energy and momentum relaxation due to electron--electron and electron--phonon scattering. During the third stage, those d-holes that relaxed to states with momenta $p<p_{F}$, where $p_{F}$ is the s-band Fermi momentum, recombine with upper-lying s-band electrons via vertical (in momentum space) transitions [see cartoon in Fig.~\ref{fig:fig1}(a)]. The overall MPL QE is given by the product $q=q_{e}q_{r}q_{d}$, where $q_{e}$, $q_{r}$ and $q_{d}$ are, respectively, the bulk QEs for d-hole excitation, relaxation, and radiative recombination stages. The latter two stages are dominated by fast nonradiative processes in noble metals \cite{apell-ps88}, including highly efficient (in the absence of energy gap between d-band and s-band) d-hole Auger scattering, which  quench d-hole momentum relaxation and radiative recombination and result in a very low bulk MPL  QE of  \cite{mooradian-prl69,apell-ps88,boyd-prb86} $q\sim 10^{-10}$. 

In plasmonic structures, QEs for individual MPL stages are independently modified. For excitation stage, a change in $q_{e}$ can be substantial if some spectral overlap between the initial excitation and surface plasmon bands is present. Here we assume, however, that the incident light frequency is either well above (for single-photon-absorption-induced MPL\cite{wilcoxon-jcp98,elsayed-cpl00,elsayed-jpcb03,feldmann-prb04,elsayed-prb05,link-jpcc11,hulst-nl11}) or well below (for multiphoton-absorption-induced MPL \cite{novotny-prb03,fourkas-nl05,wang-pnas05,bouhelier-prl05,imura-jpcb05,schuck-prl05,eichelbaum-nanotech07,durr-nl07,wang-nl07,potma-jpcc08,park-oe08,imura-jpcc09,wang-oe09,biagioni-prb09,loumaigne-nl10,wissert-nl10,biagioni-nl12}) the plasmon band so that $q_{e}$ remains unaffected. The QE for fast relaxation stage, $q_{r}$, is not directly affected by plasmons although the increased d-hole surface scattering may lead to faster momentum relaxation in small nanostructures \cite{vallee-prb04}. However, $q_{r}$ is significantly enhanced for multiphoton (vs. single-photon) excitation. Indeed,  the multiphoton d-hole excitation in noble metals is  dominated by the following incoherent sequential process \cite{imura-jpcb05,imura-jpcc09,biagioni-prb09,biagioni-nl12}: an incident photon creates a vacancy in the s-band  below  the Fermi level that is filled by a d-band electron promoted by a subsequent photon. In this way, d-holes are created with momenta predominantly below $p_{F}$ and hence need not undergo large momentum transfer before radiative recombination. In the absence of phase space restriction, $q_{r}$ drastically increases resulting in a highly efficient multiphoton-absorption-induced MPL \cite{novotny-prb03,fourkas-nl05,wang-pnas05,bouhelier-prl05,imura-jpcb05,schuck-prl05,eichelbaum-nanotech07,durr-nl07,wang-nl07,potma-jpcc08,park-oe08,imura-jpcc09,wang-oe09,biagioni-prb09,loumaigne-nl10,wissert-nl10,biagioni-nl12}. A quantitative description of the above MPL dynamics is beyond the scope of our paper. Here, we focus solely on the third (recombination) MPL stage, characterized by quantum efficiency $Q_{d}$ that is strongly influenced by plasmons, and derive the QE enhancement factor, $M=Q_{d}/q_{d}$. 

We start with excitation of Auger plasmons by a localized d-hole in a spherical metal NP of radius $a$. The decay rate of a photoexcited d-hole is given by $\Gamma_{d}=2\,{\rm Im}\Sigma_{d}$, where $\Sigma_{d}$ is d-hole self-energy due to electron--electron and electron--phonon interactions as well as interaction with light. In plasmonic nanostructures, the interactions between electrons (and holes) are described by dynamically-screened Coulomb potential  $U(\omega;{\bf r},{\bf r}')=U_{0}({\bf r}-{\bf r}')+ \delta U(\omega;{\bf r},{\bf r}')$, where $U_{0}=e^{2}/\left (\epsilon(\omega)|{\bf r}-{\bf r}'|\right )$ is the  bulk Coulomb potential ($e$ is the electron charge) and $\delta U$ contains contribution from surface excitations. In spherical NP, $\delta U$ takes the form \cite{shahbazyan-prl98,shahbazyan-prb99,shahbazyan-cp00}
\begin{equation}
\label{image}
\delta U(\omega;{\bf r},{\bf r}')=\sum_{lm}\frac{4\pi}{2l+1}\frac{e^{2}}{a^{2l+1}}\,\frac{1}{\tilde{\epsilon}_{l}(\omega)}\,\phi_{lm}({\bf r})\phi_{lm}^{\ast}({\bf r}'),
\end{equation}
where $\phi_{lm}({\bf r})=r^{l}Y_{lm}(\hat{\bf r})$ ($Y_{lm}$ are spherical harmonics) and the factor $1/\tilde{\epsilon}_{l}=(2l+1)/[l\epsilon+(l+1)\epsilon_{0}]-1/\epsilon$ contains a pole corresponding to surface plasmon with angular momentum $l$. The full decay rate $\Gamma_{d}$ can be split into sum $\Gamma_{d}=\gamma_{d}^{nr}+\Gamma_{d}^{r}+\Gamma_{d}^{p}$, where $\gamma_{d}^{nr}$ is  bulk nonradiative decay rate, $\Gamma_{d}^{r}$ is  plasmon-enhanced  radiative decay rate, and $\Gamma_{d}^{p}$
is plasmon-assisted nonradiative decay rate. The latter is determined by the contribution to  d-hole self-energy due to interband transitions under potential $\delta U$,
\begin{equation}
\Sigma_{d}^{p}=\sum_{lm}\frac{4\pi}{2l+1}\,\frac{\left |M_{lm}^{sd}\right |^{2}}{a^{2l+1}\tilde{\epsilon}_{l}(\omega)},
\end{equation}
where $M_{lm}^{sd}=e\langle s|\phi_{lm}({\bf r})|d\rangle$ is the interband transition matrix element. To evaluate $M_{lm}^{sd}$, we adopt the tight-binding approach used for calculation of local field corrections to the dielectric function \cite{wiser-pr63,nagel-prb75,krezin-prb95}. First, we expand the d-hole Bloch function in the Wannier basis as $u_{d{\bf k}}({\bf r})={\cal V}^{-1}\sum_{j}e^{i{\bf k}\cdot {\bf r}_{j}}\varphi_{d}({\bf r-r}_{j})$, where $\varphi_{d}({\bf r-r}_{j})$ is the localized hole wave-function for atomic position ${\bf r}_{j}$ and ${\cal V}$ is unit cell volume. Note that, for a given d-hole position, only one term effectively contributes to the sum. In this basis, the matrix element of a smooth (on atomic scale) function $\phi_{lm}({\bf r})$ between d-band and s-band states is $e\langle s|\phi_{lm}({\bf r})|dj\rangle={\bf d}_{sd}\cdot \nabla \phi_{lm}({\bf r}_{j})$, where ${\bf d}_{sd}=e\langle s|({\bf r-r}_{j})|dj\rangle$ is the interband dipole matrix element \cite{wiser-pr63,nagel-prb75,krezin-prb95}. The next step is to average $\langle|{\bf d}_{sd}\cdot \nabla \phi_{lm}({\bf r})|^{2}\rangle$ over orientations of ${\bf d}_{sd}$ assuming isotropic angular distribution: $\langle d_{sd}^{\mu}d_{sd}^{\nu}\rangle =\frac{1}{3}d_{sd}^{2}\delta_{\mu\nu}$. Finally, using the relation $\frac{4\pi}{2l+1}\sum_{m} |\nabla \phi_{lm}({\bf r})|^{2}=l(2l+1)r^{2l-2}$, we obtain 
\begin{equation}
\label{nonrad}
\Sigma_{d}^{p}=-\frac{d_{sd}^{2}}{3a^{3}}\sum_{l}l\left (2l+1\right )\frac{1}{\tilde{\epsilon}_{l}(\omega)}\left (\frac{r}{a}\right )^{2l-2}.
\end{equation}
The surface-induced decay rate $\Gamma_{d}^{p}=2\,\text{Im}\Sigma_{d}^{p}$ peaks at frequencies satisfying $\epsilon(\omega)+(1+1/l)\epsilon_{0}=0$ indicating Auger plasmon excitation.  A d-hole at the NP center ($r=0$) decays into dipole plasmon only with the rate \cite{shahbazyan-prl98,shahbazyan-prb99,shahbazyan-cp00} 
$\Gamma_{d1}^{p}=-2\left (d_{sd}^{2}/a^{3}\right )\text{Im}\left [1/\tilde{\epsilon}_{1}(\omega)\right ]$. In Fig.~\ref{fig:fig1}(b), we plot  $\Gamma_{d1}^{p}$ to illustrate the huge (up to $10^{5}$) difference between plasmon-assisted radiative and nonradiative  decay rates. 

Turning to MPL QE, the radiated energy density  $S(\omega)$ of a dipole with excitation energy $E^{sd}$ is given by \cite{novotny-book,pustovit-prl09,pustovit-prb10} 
\begin{equation}
\label{S}
S(\omega)\propto \frac{\omega\Gamma_{d}^{r}}{|\omega-E^{sd}+\Sigma_{d} |^{2}}=\text{Im}\,\frac{2\omega Q_{d}}{E^{sd}-\Sigma_{d}-\omega},
\end{equation}
where $\Sigma_{d}$ is d-hole full self-energy and $Q_{d}=\Gamma_{d}^{r}/\Gamma_{d}$ is  QE for d-hole radiative recombination. Summing up Eq.~(\ref{S}) over empty d-band and occupied s-band states and returning to the momentum space \cite{wiser-pr63,nagel-prb75,krezin-prb95}, we obtain 
\begin{equation}
\label{S-final}
S(\omega)\propto \omega Q_{d}(\omega)\,\tilde{\chi}''_{d}(\omega),
\end{equation}
where 
\begin{equation}
\label{chi}
\tilde{\chi}_{d}(\omega) = \frac{e^{2}}{m^{2}}\int\!\frac{d {\bf p}}{(2\pi)^{3}}
\frac{\mu^{2}}{(E_{p}^{sd})^{2}}\frac{f(E_{p}^{s})\left [1-f(E_{p}^{d})\right ]}{E_{p}^{sd}-\Sigma_{d}(\omega)-\omega} ,
\end{equation}
is interband susceptibility modified by the surface, $E_{p}^{sd}=E_{p}^{s}-E_{p}^{d}$ is vertical transition energy between d-band and s-band with dispersions $E_{p}^{d}$ and $E_{p}^{s}$, respectively, $\mu=imE_p^{sd}d_{sd}/e$ is the interband momentum matrix element ($m$ is electron mass), and $f$ is the Fermi function. Importantly, the self-energy $\Sigma_{d}$ in Eq.~(\ref{chi}) depends on $\omega$ rather than on  $E_{p}^{cd}$ \cite{shahbazyan-prl98,shahbazyan-prb99,shahbazyan-cp00}, which allows us to factor out $Q_{d}(\omega)$ in Eq.~(\ref{S-final}). The bulk radiated energy density is recovered from Eqs. (\ref{S-final}) and (\ref{chi}) by replacing $\Sigma_{d}$ with $i\gamma_{d}=i(\gamma_{nr}+\gamma_{r})$, yielding $S_{0}\propto \omega q_{d}\,\chi''_{d}(\omega)$, where $q_{d}=\gamma_{d}^{r}/(\gamma_{d}^{r}+\gamma_{d}^{nr})$ and $\chi_{d}(\omega)$ are, respectively, bulk QE and interband susceptibility. Note that in Eq.~(\ref{chi}), the plasmon contribution to $\Sigma_{d}$ plays no significant role due to the integration over off-resonant ${\bf p}$-states, so that $\tilde{\chi}_{d}(\omega)$ in Eq.~(\ref{S-final}) can be replaced by its bulk counterpart $\chi_{d}(\omega)$. The local enhancement factor, $A(\omega,{\bf r})=S/S_{0}$, then takes the form
\begin{equation}
\label{enh-qe}
A(\omega,{\bf r})=\frac{Q_{d}}{q_{d}}=\frac{R}{q_{d}\left (P+R-1\right )+1},
\end{equation}
where $R(\omega,{\bf r})=\Gamma_{d}^{r}/\gamma_{d}^{r}$ and  $P(\omega,{\bf r})=\Gamma_{d}^{p}/\gamma_{d}^{r}$  both, in general, depend  on system geometry and d-hole position.

The average enhancement factor, $M$, is obtained by averaging the local enhancement factor $A(\omega,{\bf r})$ over d-hole positions inside a nanostructure.  Here we note that d-holes \textit{spatial} distribution stays largely unaffected  during the fast relaxation stage, so we can adopt the normalized distribution function $F(\omega_{i},{\bf r})=S(\omega_{i},{\bf r})/\int d{\bf r} S$, where $S(\omega_{i},{\bf r})$ is the absorbed power density, 
\begin{equation}
S(\omega_{i},{\bf r})= \frac{\omega_{i}}{2}\epsilon''(\omega_{i})\left |{\bf E}(\omega_{i},{\bf r})\right |^{2},
\end{equation}
and ${\bf E}(\omega_{i},{\bf r})$ is the local field at incident light frequency $\omega_{i}$. Then the enhancement factor \textit{per radiating d-hole} is given by the weighted average of $A(\omega,{\bf r})$ over metal volume
\begin{equation}
\label{enh-per}
M(\omega_{i},\omega)=\int dV F(\omega_{i},{\bf r})A(\omega,{\bf r}).
\end{equation}
Equations (\ref{enh-qe})-(\ref{enh-per}) represent our model of plasmon-assisted MPL that applies to any metal structure.  Below we consider MPL from a spherical NP integrated over emission angles. Since the excitation and radiation stages are separated by the relaxation stage, the incident and emitted light polarizations are uncorrelated. In this case, $R$ and $P$ in Eq.~(\ref{enh-qe}) depend only on radial coordinate $r$, and  $F(\omega_{i},{\bf r})$ in Eq. (\ref{enh-per}) can be replaced by its average, $\bar{F}(\omega_{i},r)=\frac{1}{4\pi}\int d\Omega F(\omega_{i},{\bf r})$. Using Mie theory formulae for the electric fields inside a metal sphere \cite{bohren-book}, we have $\bar{F}(\omega_{i},r)=F_{0}(\omega_{i})R(\omega_{i},r)$ with $F_{0}=\epsilon'' k_{i}/\epsilon_{0}C_{ab}$, where $C_{ab}$ is NP absorption crosssection, $k_{i}=\omega_{i}\sqrt{\epsilon_{0}}/c$ is the wavevector and $c$ is speed of light, yielding

\begin{equation}
\label{enh-np}
M=\frac{\epsilon''(\omega_{i}) k_{i}}{C_{ab}(\omega_{i})\epsilon_{0}}\int  \frac{dV\, R(\omega_{i},r)R(\omega,r)}{q_{d}\left [P(\omega,r)+R(\omega,r)-1\right ]+1}.
\end{equation}
For small NP ($a\ll \lambda$), $R$ and $P$ are derived, respectively, from Eqs. (\ref{rad}) and (\ref{nonrad}) using  $\gamma_{d}^{r}=4d_{sd}^{2}\omega^{3}\sqrt{\epsilon_{0}}/3c^{3}$, and $C_{ab}=4\pi k_{i}\alpha''_{1}$; the formulas for NPs of arbitrary size \cite{bohren-book,chew-jcp87} are provided in the Supporting Information.

\section{Numerical results}

Calculations were performed for an Au NP in water using standard Au parameters \cite{cooper-pr65} along with bulk d-hole decay time of 40 fs \cite{elsayed-jpcb03,feldmann-prb04,elsayed-prb05} rendering $q_{d}$ in the interval (3--4)$\times 10^{-7}$ within the plasmon band range 500--600 nm. Angular momenta up to $l_{max}=100$ were included in the evaluation of $\Gamma_{d}^{p}$ (in small NP, $l_{max}$ is restricted by $ap_{F}$), and experimental Au dielectric function was used throughout.

%  %%%%%%%%%%%%%%%%%%%%%%%%%%%%%%%%%%
  \begin{figure}
%  \begin{center}
 % \centering
  \includegraphics[width=0.95\columnwidth]{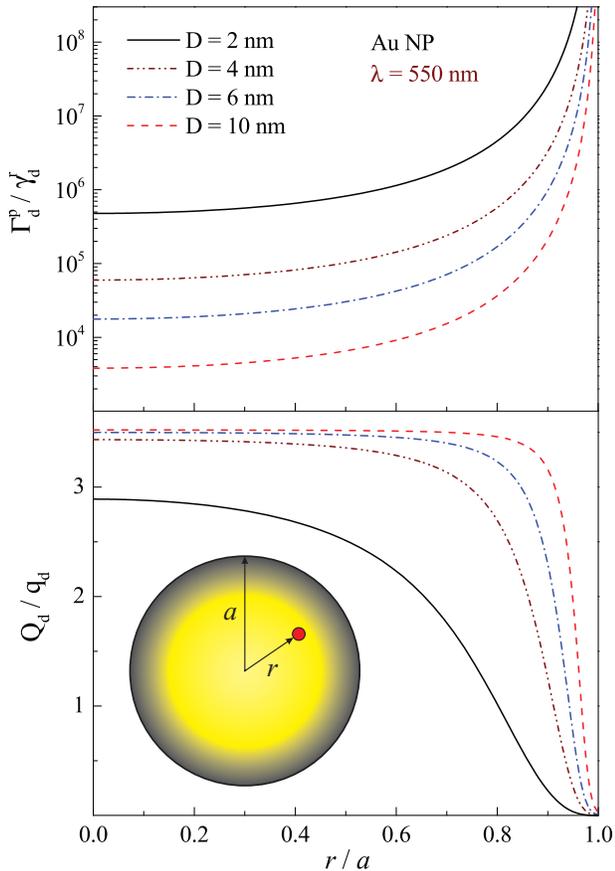}
%  \end{center}
 \caption{\label{fig:fig2} Surface-induced nonradiative decay rate (a) and local enhancement factor (b) near plasmon resonance ($\lambda=550$ nm) is shown vs. d-hole radial position in Au NP.}
  \end{figure}
%  %%%%%%%%%%%%%%%%%%%%%%%%%%%%%%%%%%

In Fig.~\ref{fig:fig2}(a), we plot $\Gamma_{d}^{p}$ normalized to $\gamma_{d}^{r}$ against d-hole position $r$ for $\omega$ near the dipole plasmon resonance (compare to Fig.~\ref{fig:fig1}). For d-hole at the NP center, only the dipole Auger plasmon is excited, but for $r\sim a$, $\Gamma_{d}^{p}$  increases by several orders of magnitude due to the high-$l$ contribution [see Eq.~(\ref{nonrad})].  Near NP boundary, $\Gamma_{d}^{p}$ exceeds significantly the bulk d-hole decay rate $\gamma_{d}^{nr}$, leading to drastic reduction of local enhancement factor $A(\omega,{\bf r})$ [see Fig.~\ref{fig:fig2}(b)]. This effect is especially strong in small NP ($D=2$ nm) where $\Gamma_{d}^{p}$ and $\gamma_{d}^{nr}$ are largely comparable in most of NP volume (i.e., away from NP center and its boundary). Note that for $D$ smaller than the skin depth ($\approx 20$ nm for Au), d-holes are distributed nearly uniformly throughout NP volume and their distribution function is simply $F=3/4\pi a^{3}$. In this case, it is NP central region that mainly contributes to MPL, i.e., NP has a "bright" core and "dark" edge that expands toward NP center as its size decreases. 

%  %%%%%%%%%%%%%%%%%%%%%%%%%%%%%%%%%%
  \begin{figure}[bt]
  \begin{center}
%  \centering
  \includegraphics[width=0.9\columnwidth]{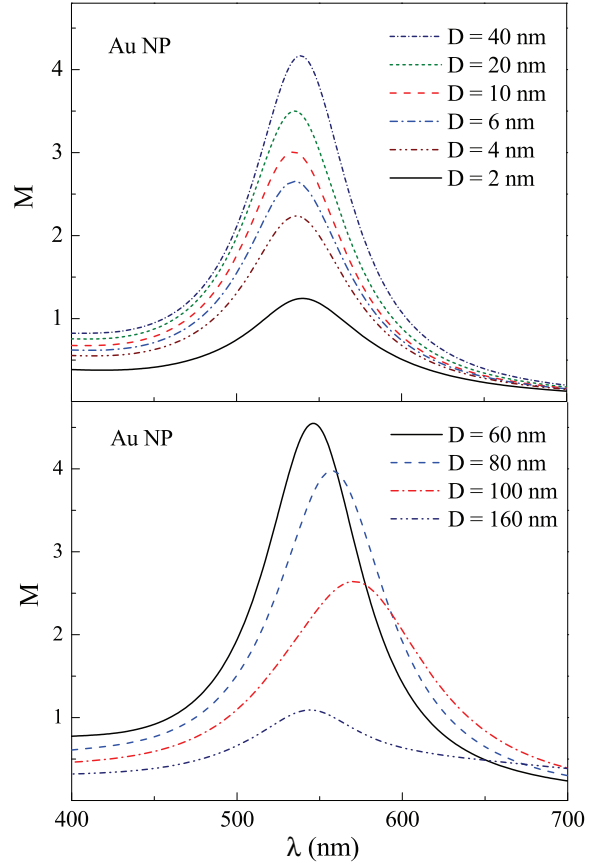}
  \end{center}
\caption{\label{fig:fig3} MPL enhancement factor for Au NP with diameters in the range 2--40 nm (a) and 60--160 nm (b). The maximal enhancement is achieved for $D=60$ nm NP.}
  \end{figure}
%  %%%%%%%%%%%%%%%%%%%%%%%%%%%%%%%%%%

In Fig.~\ref{fig:fig3} we plot the average enhancement factor $M$, calculated from Eq.~(\ref{enh-np}), for Au NPs with diameters between 2 nm and 160 nm. In a wide size range of 4--100 nm, $M$ varies only weakly (within factor of 2), and the strongest enhancement is achieved for $D=60$ nm NP, consistent with experiment \cite{fourkas-nl05}. For small NP ($D=2$ nm),  M falls by a factor $\sim$4 relative to maximum enhancement, also consistent with the data \cite{feldmann-prb04,fourkas-nl05}. Note that this remarkable agreement with experiment is achieved without use of fitting parameters. For larger NPs ($D=80$ and 100 nm), the MPL peak is redshifted \cite{novotny-prb03}, and with a further size  increase  ($D=160$ nm), the plasmon band nearly disappears and MPL is diminished. Thus, the emergence of optimal NP sizes in the range of 40--80 nm is due to the interplay between two different damping mechanisms: weakening of plasmon resonance in larger NPs due to electromagnetic effects (retardation) and MPL quenching in smaller NPs due to quantum effects (excitation of Auger plasmons).

\section{Conclusion}
In summary, we presented a microscopic theory for plasmon-enhanced metal photoluminescence. We have shown that,  in small nanostructures, plasmon-enhanced luminescence is hindered due to highly efficient excitation of Auger plasmons by photoexcited d-holes. Our numerical calculations of the MPL enhancement factor for spherical Au nanoparticles are in excellent agreement with experiment. 

This work was supported in part by NSF Grants DMR-1206975 and HRD-0833178.

%%%%%%%%%%%%%%%%%%%%%%%%%%%%%%%%%%%%%%

\section*{Supporting Information}

Here we present expressions for the surface-induced decay of an excited dipole comprised of a localized d-hole and s-electron  inside a spherical NP of radius $a$. This problem is mathematically similar to the decay of a dipole placed inside a dielectric sphere \cite{chew-jcp87}, so we adopt that solution here while noting the differences between dielectric and metal spheres. The decay rate is related to the electric field Green dyadic, ${\bf G}(\omega;{\bf r},{\bf r}')$, at the dipole position ${\bf r}$ as \cite{novotny-book}
\begin{equation}
\label{rate-green}
\Gamma_{d}=\frac{8\pi\omega^{2}}{c^{2}}\,\text{Im}\,{\bf d}\cdot {\bf G}(\omega;{\bf r},{\bf r})\cdot {\bf d},
\end{equation}
where ${\bf d}=d {\bf n}$ is the dipole moment and averaging over its orientations ${\bf n}$ is implied (the effects of electron-electron and electron-phonon scattering can be incorporated by adding the bulk decay rate  $\gamma_{d}^{nr}$, see main text). The Green dyadic has free and scattered parts, ${\bf G}={\bf G}^{0}+{\bf G}^{s}$, where $G_{\mu\nu}^{0}(\omega;{\bf r},{\bf r})=i\delta_{\mu\nu}k/6\pi$ (here $k=\omega\sqrt{\epsilon}/c$), and ${\bf G}^{s}$ has the form \cite{chew-jcp87}
\begin{align}
&{\bf G}^{s}(\omega;{\bf r},{\bf r})=\frac{ik}{4\pi}\sum_{l}(2l+1)\Biggl [E_{l}l(l+1)\left [\frac{j_{l}(x)}{x}\right ]^{2}\hat{\bf r}\hat{\bf r}
~~~
\nonumber\\
&~~~~
+\frac{1}{2}\left [E_{l}\left [\frac{[xj_{l}(x)]'}{x}\right ]^{2}+F_{l}j_{l}^{2}(x)\right ]\left (\hat{\bm \theta}\hat{\bm \theta}+\hat{\bm \phi}\hat{\bm \phi}\right )\Biggr],
\end{align}
where $(\hat{\bf r},\hat{\bs \theta},\hat{\bs \phi})$ are unit vectors in spherical coordinates, $j_{l}(x)$ is the spherical Bessel function and $x=kr$. The Mie coefficients $E_{l}$ and $F_{l}$ are given by
\begin{align}
\label{EF}
&E_{l}=\left [\epsilon_{0}h_{l}(x_{2})\left [x_{1}h_{l}(x_{1})\right ]'-\epsilon h_{l}(x_{1})\left [x_{2}h_{l}(x_{2})\right ]'\right ]/D_{l}^{e},
\nonumber\\
&F_{l}=\left [h_{l}(x_{2})\left [x_{1}h_{l}(x_{1})\right ]'- h_{l}(x_{1})\left [x_{2}h_{l}(x_{2})\right ]'\right ]/D_{l}^{m},
\end{align}
with
\begin{align}
\label{DD}
&D_{l}^{e}=\epsilon j_{l}(x_{1})\left [x_{2}h_{l}(x_{2})\right ]'-\epsilon_{0} h_{l}(x_{2})\left [x_{1}j_{l}(x_{1})\right ]',
\nonumber\\
&D_{l}^{m}=j_{l}(x_{1})\left [x_{2}h_{l}(x_{2})\right ]'- h_{l}(x_{2})\left [x_{1}j_{l}(x_{1})\right ]',
\end{align}
where $h_{l}(x)=j_{l}(x)+iy_{l}(x)$ is the spherical Hankel function and we adopted the notations $x_{1}=\left (\omega\sqrt{\epsilon}/c\right )a$ and $x_{2}=\left (\omega\sqrt{\epsilon_{0}}/c\right )a$ (prime stands for derivative). Normalizing $\Gamma_{d}$ by radiative decay rate in outside medium, $\gamma_{d}^{r}=4d^{2}\omega^{3}\sqrt{\epsilon_{0}}/3c^{3}$, and averaging over dipole orientations, we obtain
\begin{align}
\label{rate-full}
&\frac{\Gamma_{d}}{\gamma_{d}^{r}}=\text{Re}\,\sqrt{\frac{\epsilon}{\epsilon_{0}}}\Biggl [1+\sum_{l}\left (l+\frac{1}{2}\right )\biggl [E_{l}l(l+1)\left [\frac{j_{l}(x)}{x}\right ]^{2}
\nonumber\\
&
~~~~~~~~~
+E_{l}\biggl [\frac{[xj_{l}(x)]'}{x}\biggr]^{2}+F_{l}j_{l}^{2}(x)\biggr]\Biggr ].
\end{align}
The radiative decay rate $\Gamma_{d}^{r}$ is derived by integrating the Poynting vector over a sphere with infinite radius \cite{chew-jcp87},
\begin{align}
\label{rate-rad}
&\frac{\Gamma_{d}^{r}}{\gamma_{d}^{r}}=\sum_{l}\left (l+\frac{1}{2}\right )\Biggl [\left |\frac{\epsilon}{ka D_{l}^{e}}\right |^{2}\Biggl [l(l+1)\left |\frac{j_{l}(kr)}{kr}\right |^{2}
\nonumber\\
&~~~~~~~~~~~~~~
+\biggl |\frac{[krj_{l}(kr)]'}{kr}\biggr|^{2}\Biggr]
+\biggl |\sqrt{\frac{\epsilon}{\epsilon_{0}}}\frac{j_{l}(kr)}{kaD_{l}^{m}}\biggr |^{2}\Biggr],
\end{align}
with coefficients $D_{l}^{e}$ and $D_{l}^{m}$ given by Eq.~(\ref{DD}). Note that for dielectric sphere ($\epsilon>0$), the two rates (\ref{rate-full}) and (\ref{rate-rad}) coincide \cite{chew-jcp87}. For metal sphere with complex dielectric function $\epsilon(\omega)$ considered here, radiative rate is still given by Eq.~(\ref{rate-rad}), while now $\Gamma_{d}=\Gamma_{d}^{p}+\Gamma_{d}^{r}$.  In our numerical calculations, we used the above normalized rates, 
$R(\omega,{\bf r})=\Gamma_{d}^{r}/\gamma_{d}^{r}$ and $P(\omega,{\bf r})=\Gamma_{d}^{p}/\gamma_{d}^{r}$. In the longwave limit, $ak\ll 1$, it can be readily checked that those coincide with the corresponding expressions  in the main text.

Finally, the  NP absorption crossection can be found as the difference between the extinction and scattering crossections, $C_{ab}=C_{ex}-C_{sc}$, defined in the standard way \cite{bohren-book}
\begin{align}
\label{cross}
&C_{ex}=\frac{2\pi}{k_{0}^{2}}\text{Re}\sum_{l}\left (2l+1\right )\left (a_{l}+ b_{l}\right ) ,
\nonumber\\
&C_{sc}=\frac{2\pi}{k_{0}^{2}}\text{Re}\sum_{l}\left (2l+1\right )\left (|a_{l}|^{2}+ |b_{l}|^{2}\right ),
\end{align}
where $A_{l}$ and $B_{l}$ are the Mie coefficients,
\begin{align}
\label{AB}
&A_{l}=\left [\epsilon j_{l}(x_{1})\left [x_{2}j_{l}(x_{2})\right ]'-\epsilon_{0}j_{l}(x_{2})\left [x_{1}j_{l}(x_{1})\right ]'\right ]/D_{l}^{e},
\nonumber\\
&B_{l}=\left [j_{l}(x_{1})\left [x_{2}j_{l}(x_{2})\right ]'-j_{l}(x_{2})\left [x_{1}j_{l}(x_{1})\right ]'\right ]/D_{l}^{m}.
\end{align}
These expressions were used for the calculation of d-hole distribution function.

{}

\end{document}